\documentclass[11pt]{article}

\usepackage[utf8]{inputenc}
\usepackage[T1]{fontenc}
\usepackage{amsmath, amssymb, amsthm}
\usepackage{newtxtext}
\usepackage{newtxmath}
\usepackage{comment}
\usepackage[margin=1in]{geometry}

\numberwithin{equation}{section}

\theoremstyle{plain}

\theoremstyle{definition}

\usepackage{graphicx}
\usepackage{subcaption}
\usepackage{booktabs}
\usepackage[font=small, labelfont=bf]{caption}
\usepackage{float}
\graphicspath{{figures/}}
\usepackage{booktabs}
\usepackage{multirow}

\usepackage[numbers]{natbib}
\usepackage[hidelinks]{hyperref}
\hypersetup{
    pdftitle={Your Title},
    pdfauthor={Your Name},
}

\usepackage{enumitem}
\usepackage{microtype}
\usepackage{xcolor}
\usepackage{comment}

\title{Diffusion-Based Synthesis of 3D T1w MPRAGE Images from Multi-Echo GRE with Multi-Parametric MRI Integration \\[4pt]}

\author{
  Sizhe Fang\thanks{Department of Radiology and Imaging Sciences, Emory University, Atlanta, GA, USA. 
  School of Biological Sciences, Georgia Institute of Technology, Atlanta, GA, USA. 
  Email: sfang86@gatech.edu}
  \and 
  Deqiang Qiu\thanks{Corresponding author. Department of Radiology and Imaging Sciences, 
  Emory University, Atlanta, GA, USA. Email: dqiu3@emory.edu}
}
\date{November 2025}

\begin{document}
\maketitle

\begin{abstract}
Multi-echo Gradient Echo (mGRE) sequences provide valuable quantitative parametric maps, such as Quantitative Susceptibility Mapping (QSM) and transverse relaxation rate (R2*), which are sensitive to tissue iron and myelin. However, structural morphometry typically relies on separate T1-weighted MPRAGE acquisitions, prolonging scan times. We propose a deep learning framework to synthesize high-contrast 3D T1w MPRAGE images directly from mGRE data, thereby streamlining neuroimaging protocols. 
We developed a novel multi-parametric conditional diffusion model based on the Fast-DDPM architecture. Unlike conventional intensity-based synthesis, our approach integrates iron-sensitive QSM and R2* maps as physical priors to address the challenge of contrast ambiguity in iron-rich deep gray matter structures. We trained and validated the model on a cohort of 175 healthy subjects. Performance was evaluated against established U-Net and GAN-based baselines using perceptual metrics and downstream segmentation accuracy. Uniquely, we assessed the biological plausibility of the synthesized images by replicating population-level statistical associations with age and sex.
The proposed framework significantly outperformed baseline methods, achieving superior perceptual quality and segmentation accuracy, particularly in subcortical regions like the thalamus and pallidum. Crucially, the synthesized images preserved essential biological dependencies: regression analyses demonstrated high concordance in age-related atrophy rates, aging effect sizes, and sexual dimorphism patterns compared to ground truth data.
By effectively leveraging quantitative MRI priors, our diffusion-based method generates strictly biologically plausible T1w images suitable for reliable clinical morphometric analysis. This approach offers a promising pathway to reduce acquisition time by deriving structural contrasts retrospectively from quantitative mGRE sequences.

\end{abstract}

\section{Introduction}
Magnetic Resonance Imaging (MRI) provides essential anatomical and functional information for clinical diagnosis and neuroimaging research through diverse imaging sequences. T1-weighted Magnetization Prepared Rapid Gradient Echo (T1w MPRAGE) images are widely acquired in neuroimaging studies due to their superior gray matter-white matter contrast, making them essential for brain morphometry and tissue segmentation \cite{mugler1990three, reuter2012within}. Multi-echo Gradient Echo (mGRE) sequences, on the other hand, enable quantitative parametric mapping including transverse relaxation rate (R2*) and quantitative susceptibility mapping (QSM), which are highly sensitive to iron deposition and myelin content in brain tissue \cite{barbosa2015quantifying, liu2015quantitative}. In typical neuroimaging protocols, both sequences are acquired to provide complementary structural and quantitative information. However, this results in prolonged scan times, increasing patient burden and scanning costs. The ability to synthesize MPRAGE images from mGRE data could streamline imaging workflows while maintaining access to both structural morphometry and quantitative parametric maps.

Medical image synthesis using deep learning has shown promise in addressing the limitations of multi-sequence acquisition protocols \cite{wang2021review, kazerouni2023diffusion}. Early approaches relied on traditional machine learning methods such as atlas-based techniques and random forests, which suffered from limited generalizability and often required manual intervention \cite{johnstone2018systematic, roy2013atlas}. The advent of deep learning brought significant advances, with convolutional neural networks (CNNs) and U-Net architectures demonstrating success in various cross-modality synthesis tasks \cite{ronneberger2015unet, han2017mr}. Generative Adversarial Networks (GANs) further improved synthesis quality through adversarial training, becoming widely adopted for medical image generation \cite{goodfellow2014generative, isola2017image}. However, GANs can suffer from training instability and mode collapse, which may compromise the reliability of synthesized medical images \cite{skandarani2023gans}.

Recently, denoising diffusion probabilistic models (DDPMs), or diffusion models, have emerged as a powerful alternative for medical image synthesis. Unlike GANs that rely on adversarial training, diffusion models generate images through an iterative denoising process, offering superior stability and sample quality \cite{ho2020denoising, shi2025diffusion}. Recent medical imaging applications have demonstrated their effectiveness in cross-modality MRI synthesis and image reconstruction \cite{jiang2023coladiff, khader2023denoising}. Notably, the Fast-DDPM framework has made diffusion models more practical by reducing sampling steps from 1,000 to 10 while maintaining generation quality \cite{jiang2025fastddpm}.

For quantitative neuroimaging studies using mGRE images, MPRAGE acquisitions serve a critical role by providing anatomical references for tissue segmentation and region-based analysis of QSM and R2* maps. QSM quantifies magnetic susceptibility distributions within tissues and has shown particular value in detecting iron accumulation in deep gray matter structures, which is a biomarker relevant to aging and neurodegenerative diseases \cite{langkammer2012quantitative, acosta2016vivo}. R2* mapping similarly reflects tissue microstructural properties influenced by iron content and field inhomogeneities \cite{cohen2014learn}. While automated segmentation tools like FreeSurfer require high-quality T1-weighted images for accurate parcellation, acquiring separate MPRAGE scans adds 5-6 minutes to imaging protocols. Ryu et al. previously demonstrated the feasibility of synthesizing MPRAGE from mGRE using a 3D U-Net, achieving good segmentation accuracy and correlation for susceptibility measurements \cite{ryu2019synthesizing}. However, intensity-based approaches may struggle to resolve contrast ambiguity in iron-rich deep gray matter structures. Building upon this foundation, recent advances in generative modeling and the availability of derived parametric maps from mGRE data motivate further exploration of this synthesis task.

In this study, we investigate the application of conditional diffusion models for synthesizing T1w MPRAGE images from mGRE data with multi-parametric MRI integration. Specifically, we adapt the Fast-DDPM framework and explore whether incorporating derived QSM and R2* maps alongside mGRE magnitude images can enhance synthesis quality. We systematically evaluate our approach through both qualitative and quantitative methods. Comparative experiments with U-Net and GAN baselines assess the relative advantages of the diffusion-based approach. Finally, we demonstrate the practical utility of synthesized MPRAGE images by applying them to investigate age-related brain structural changes. This work represents the first application of diffusion models to mGRE-to-MPRAGE synthesis and provides insights into multi-parametric conditioning strategies for cross-contrast MRI synthesis.

\section{Materials and Method}
\label{sec:method}

\subsection{Subjects}
This study included 175 participants from the Emory Brain Imaging Program. The cohort comprised 129 females (mean age ± SD = 54.49 ± 18.46 years) and 46 males (mean age ± SD = 48.85 ± 20.40 years), with an overall mean age of 53.0 ± 19.1 years. All participants provided written informed consent in accordance with a protocol approved by the Institutional Review Board of Emory University. All participants were cognitively normal individuals. Subjects with a prior history of neurological or psychiatric disorders were excluded from this study.

\subsection{MRI Acquisition}
All MRI scans were performed on a Siemens Magnetom Prisma 3T scanner equipped with a 32-channel phased-array head coil at the Brain Health Center of Emory University. A multi-echo 3D spoiled gradient echo (mGRE) sequence was acquired with the following parameters: voxel size = 0.72 × 0.72 × 1.44 mm³, repetition time (TR) = 37 ms, five echoes with echo times (TE) = 6.61, 12.85, 19.09, 25.33, and 31.57 ms, and flip angle (FA) = 15°. Three-dimensional T1-weighted (T1w) images were acquired using a magnetization-prepared rapid gradient-echo (MPRAGE) sequence with the following parameters: TR = 2300 ms, TE = 2.96 ms, inversion time (TI) = 900 ms, FA = 9°, 208 sagittal slices with 1 mm slice thickness, in-plane matrix size = 256 × 240, and isotropic voxel resolution of 1 × 1 × 1 mm³.

\subsection{Data Pre-Processing}
QSM images were reconstructed from the mGRE data using an in-house processing pipeline\cite{qiu2014mr}, which included phase unwrapping, background field removal, and magnetic field-to-susceptibility inversion. R2* maps were calculated by performing monoexponential fitting to the magnitude signals across the five echoes of the mGRE sequence (FA = 15°). Detailed descriptions of the QSM and R2* reconstruction procedures have been reported in our previous publications\cite{lin2022brain,lin2023magnetic}. T1w MPRAGE images were padded and cropped to a final matrix size of 256 × 256 × 160 voxels. The first echo of the mGRE sequence was rigidly registered to the 1 mm isotropic MPRAGE images using the FLIRT toolbox\cite{jenkinson2012fsl} within FSL (FMRIB Software Library, Oxford, UK), and the resulting transformation matrix was applied to register the QSM and R2* maps to the MPRAGE space.

Following registration, all images were independently normalized to a range of -1 to 1 for each modality group. A brain mask extracted from the registered mGRE images was applied to remove extracerebral tissues from the T1w MPRAGE images. Given that brain iron is predominantly concentrated in basal ganglia and other deep gray matter structures\cite{hallgren1958effect,madden2023quantitative,ghadery2015r2mapping}, anatomical masks derived from the MNI152 standard template (registered to individual MPRAGE space) were applied to the QSM and R2* maps to extract five bilateral regions of interest: Caudate, Putamen, Pallidus, Accumbens, and Thalamus. To ensure complete coverage of these structures, the masks were dilated by 3 voxels. All voxels outside the masked regions were set to background values (-1 for QSM). Finally, each 3D volume was converted into 2D axial slices with dimensions of 256 × 256 pixels to accommodate the input format of the diffusion model.

\subsection{Model Implement and Trianing Details}
{The dataset was partitioned into a training set (n = 77; 56 females, 21 males; mean age = 53.22 ± 19.13 years) and a test set (n = 98; 73 females, 25 males; mean age
   = 52.84 ± 19.17 years). No significant differences were observed between the two sets in terms of sex distribution (p = 0.928) or age (p = 0.896).

  We implemented a conditional diffusion model based on the Fast-DDPM framework \cite{jiang2025fastddpm} to synthesize T1w MPRAGE images from multi-echo GRE images. The overall
  architecture is illustrated in Fig \ref{fig:architecture}. The network consists of an encoder-decoder structure with skip connections. To accommodate multi-channel inputs, the initial
   convolutional layer was modified to accept N+1 channels, where N represents the number of input modalities and the additional channel corresponds to the noisy
  target during forward diffusion. Two model configurations were trained: the first used five mGRE magnitude images as input (N=5), while the second incorporated QSM
   and R2* maps in addition to the five mGRE images (N=7).

  The encoder employs convolutional blocks with a base channel dimension of 128 and channel multipliers of [1, 1, 2, 2, 4, 4]. Each level contains two residual blocks
  with group normalization (32 groups) and Swish activation. Self-attention layers are applied at the 16×16 resolution to capture long-range spatial dependencies. The
  decoder mirrors the encoder structure with transposed convolutions for upsampling. Skip connections transfer features from encoder to decoder at corresponding
  resolution levels.

  Following the Fast-DDPM methodology, we reduced the number of diffusion timesteps from 1000 to 100. The forward diffusion process employs a linear-$\beta$ noise schedule
  with $\beta_1 = 0.0001$ and $\beta_T = 0.02$ defined over 1000 virtual timesteps. During both training and inference, 100 timesteps were uniformly sampled from this
  schedule at intervals of 10 (i.e., timesteps {1, 11, 21, ..., 991}). This approach enables the model to focus its learning on the most informative noise levels
  rather than uniformly distributing capacity across all possible diffusion stages. The training objective minimizes the mean squared error between predicted and
  ground truth noise:

  $$  \mathcal{L} = \mathbb{E}_{t,x_0,\varepsilon,c} \left[ \left\| \varepsilon - \varepsilon_\theta\left(\sqrt{\bar{\alpha}_t}x_0 + \sqrt{1-\bar{\alpha}_t}\varepsilon, c,
   t\right) \right\|^2 \right]$$

  where $\varepsilon_\theta$ is the noise prediction network, $x_0$ is the ground truth T1w image, $c$ represents the multi-channel input conditions, $t$ is uniformly
  sampled from the 100 selected timesteps, $\varepsilon \sim \mathcal{N}(0, I)$ is Gaussian noise, and $\bar{\alpha}t = \prod{i=1}^t (1-\beta_i)$.

  The network was optimized using Adam \cite{kingma2014adam} with learning rate $2 \times 10^{-5}$, $\beta_1 = 0.9$, $\beta_2 = 0.999$, and batch size 16 for 1,400,000 iterations.
  Antithetic sampling was employed for timestep selection, where for each batch, half of the samples used timestep index $i$ and the other half used the complementary
  index from the opposite end of the timestep sequence. An exponential moving average (EMA) of model weights was maintained with decay rate 0.999. Model checkpoints
  were saved every 50,000 iterations. Training was performed on 4 NVIDIA A40 GPUs.

  During inference, synthetic T1w MPRAGE images were generated using the generalized denoising procedure with deterministic sampling ($\eta = 0$). Starting from Gaussian noise $x_T \sim \mathcal{N}(0, I)$, the reverse diffusion process iteratively denoises the latent representation conditioned on the input images over the same 100
  uniformly sampled timesteps used during training. The EMA model weights were used for all evaluations. The proposed method was compared against U-Net \cite{ronneberger2015unet} and Pix2Pix \cite{isola2017image} baselines trained on the same dataset.
  
  \begin{figure}[htbp]
    \centering
    \includegraphics[width=0.8\textwidth]{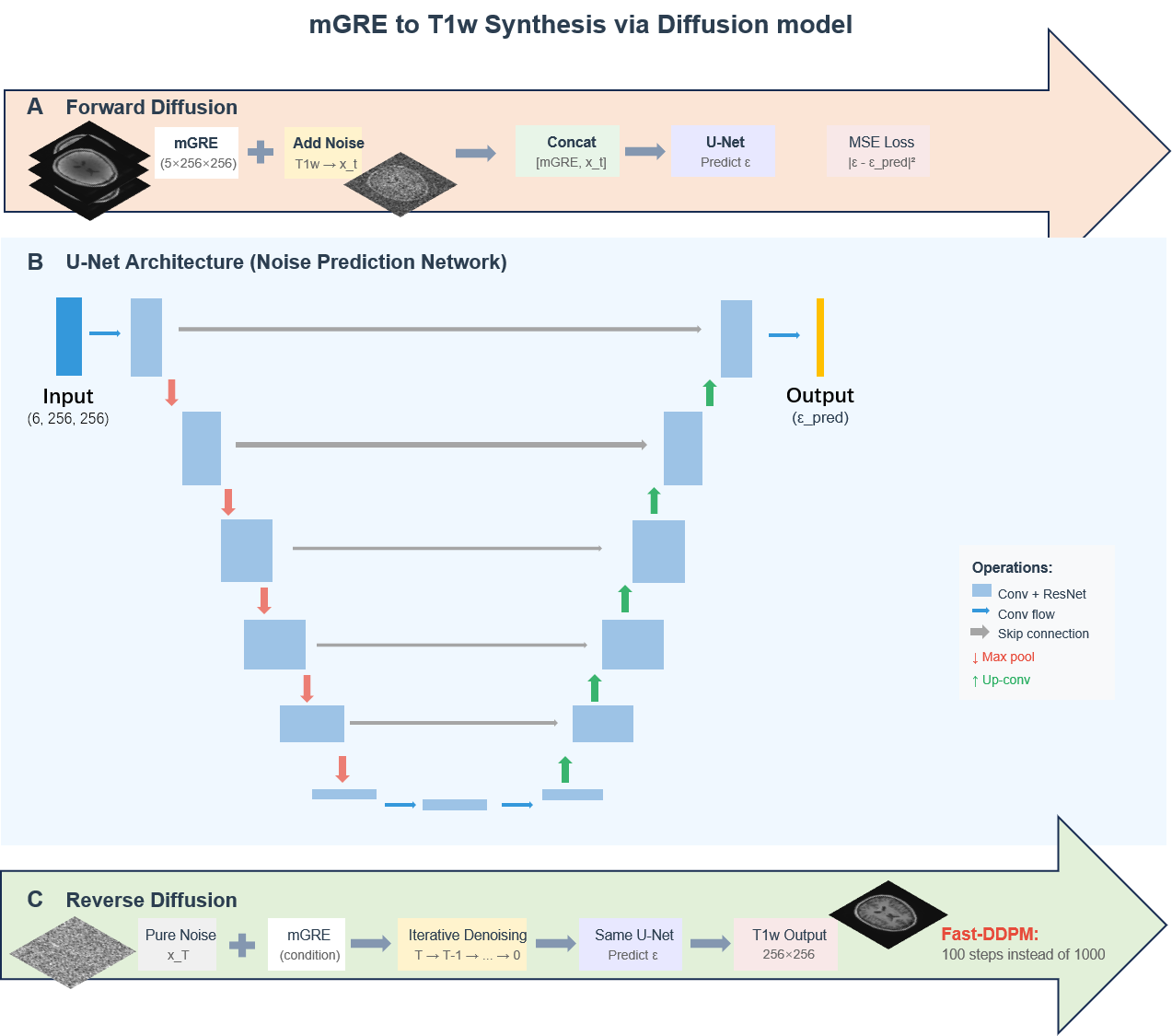}
    \caption{Architecture of the diffusion model for multi-echo gradient echo (mGRE) to T1-weighted image synthesis( 2D slice as input). (A) Forward diffusion process: During training, Gaussian noise is progressively added to the T1w target image. The U-Net learns to predict the noise at each timestep. (B) U-Net noise prediction network: A standard encoder-decoder architecture that takes as input a 6-channel image (5-channel mGRE + 1-channel noisy image) or 8-channel image( 6-channel image + 1-channel QSM + 1-channel R2*) and outputs a 1-channel predicted noise map. (C) Reverse diffusion process: During inference, the model starts from pure noise and iteratively denoises it under mGRE guidance to generate the T1w image. Fast-DDPM\cite{jiang2025fastddpm} is used to accelerate sampling compared to standard DDPM. }
    \label{fig:architecture}
\end{figure}
  }

\subsection{Evaluation Metrics}
{

  \subsubsection{Image Quality Assessment}

  In addition to qualitative analysis,to quantitatively evaluate the quality of synthesized T1w MPRAGE images, we employed two widely used metrics:
  peak signal-to-noise ratio (PSNR) and structural similarity index measure (SSIM)\cite{wang2004ssim}. PSNR measures the
  pixel-wise fidelity between the synthesized and ground truth images, defined as:

  $$\text{PSNR} = 10 \cdot \log_{10} \left( \frac{\text{MAX}_I^2}{\text{MSE}} \right)$$

  where $\text{MAX}I$ represents the maximum possible pixel value and $\text{MSE} = \frac{1}{N}\sum{i=1}^{N}(I_i -
  \hat{I}_i)^2$ is the mean squared error between the ground truth image $I$ and synthesized image $\hat{I}$ over
  $N$ pixels.

  SSIM evaluates the structural similarity between two images by comparing luminance, contrast, and structure,
  computed as:

  $$\text{SSIM}(I, \hat{I}) = \frac{(2\mu_I\mu_{\hat{I}} + C_1)(2\sigma_{I\hat{I}} + C_2)}{(\mu_I^2 +
  \mu_{\hat{I}}^2 + C_1)(\sigma_I^2 + \sigma_{\hat{I}}^2 + C_2)}$$

  where $\mu_I$ and $\mu_{\hat{I}}$ are the mean intensities, $\sigma_I^2$ and $\sigma_{\hat{I}}^2$ are the
  variances, $\sigma_{I\hat{I}}$ is the covariance, and $C_1$, $C_2$ are stabilization constants.

  \subsubsection{Segmentation-Based Functional Assessment}

  Since T1w MPRAGE images are primarily acquired for anatomical segmentation rather than quantitative analysis, we
  evaluated the functional utility of synthesized images by comparing tissue segmentation performance. Both
  synthesized and ground truth MPRAGE images were automatically segmented using FreeSurfer \cite{reuter2012within} without manual
  intervention. Segmentation accuracy was quantified using the Dice similarity coefficient (DSC) \cite{zou2004statistical}:

  $$\text{DSC}(G, S) = \frac{2|G \cap S|}{|G| + |S|}$$

  where $S$ represents the binary segmentation from synthesized MPRAGE and $G$ represents the segmentation from
  ground truth MPRAGE. DSC values range from 0 to 1, with higher values indicating better spatial overlap. We
  calculated DSC for eight representative brain regions: cortex, cerebral white matter, putamen, pallidum,
  thalamus, caudate, amygdala, and hippocampus. For bilateral structures, left and right hemispheres were
  analyzed separately and then averaged. Statistical significance of differences between methods was assessed using
   paired t-tests.

  To quantify the reliability of volumetric measurements across different segmentation models, we computed the
  intraclass correlation coefficient (ICC) with two-way mixed effects model (ICC2k) \cite{shrout1979intraclass} for the eight brain
  regions. ICC values above 0.75 indicate good reliability \cite{koo2016guideline}.
  
\begin{comment}
   For visualization of cortical segmentation quality, we performed surface rendering of DSC values across cortical
  regions using the Desikan-Killiany atlas (DK-68) \cite{desikan2006automated}.   
\end{comment}

  \subsubsection{Baseline Comparisons}

  For comparison, we trained two baseline methods using the same generator architecture: (1) supervised learning
  with L1 loss (U-Net baseline), and (2) Pix2Pix \cite{isola2017image} with combined L1 and adversarial losses. All methods were
  trained on the same dataset to ensure fair comparison.

  \subsubsection{Biological Plausibility Assessment}

  To assess whether synthesized images preserve biologically meaningful information for clinical research on aging
  effects, we performed regression analyses on brain regional volumes and cortical thickness. We compared the
  best-performing generative model against ground truth using age and sex as covariates.

  Independent multiple linear regression (MLR) analyses were conducted on nine regions of interest (ROIs) to
  evaluate whether synthesized data ($D_{\text{GEN}}$) could replicate the established associations between brain
  morphology and demographic variables observed in ground truth data ($D_{\text{GT}}$). For each ROI, the
  regression model was formulated as:

  $$Y_{\text{ROI}} = \beta_0 + \beta_{\text{Age}} \cdot \text{Age} + \beta_{\text{Sex}} \cdot \text{Sex} +
  \beta_{\text{eTIV}} \cdot \text{eTIV} + \epsilon$$

  where $Y_{\text{ROI}}$ represents regional volume or mean cortical thickness (cortical thickness measures are not influenced by head size; therefore, correction for eTIV is not required), and $\epsilon$ is the error term.
  Age and sex were included as independent variables. Estimated total intracranial volume (eTIV) was
  included as a covariate for volumetric analyses to correct for head size but excluded from cortical thickness
  models \cite{buckner2004unified}.

  The aging effect was quantified using the regression coefficient $\beta_{\text{Age}}$, which indicates the annual rate of change in the regional measure, along with its corresponding $p$-value. Furthermore, we calculated Cohen's $f$ to evaluate the effect size of aging, derived as the square root of Cohen's $f^2$ \cite{cohen1988statistical}. To evaluate the fidelity of generated data, we compared these derived metrics (adjusted $R^2$, $\beta_{\text{Age}}$, Cohen's $f$ for aging effect, and Cohen's $d$ for sex differences) between $D_{\text{GEN}}$ and $D_{\text{GT}}$ groups. High alignment between coefficients would indicate that the generative model successfully preserves the biological dependencies present in the real population.
}

\section{Results}

\subsection{Image Synthesis Performance}
We evaluated the image synthesis quality of the proposed diffusion model against two baseline methods (U-Net and
  Pix2Pix) using both quantitative metrics and qualitative visual assessment. For quantitative evaluation, slices
  containing only background (identical to ground truth) were excluded from metric calculations to avoid bias.

  \textbf{Quantitative Analysis.} Table~\ref{tab:results_comparison} presents the quantitative comparison results across
  four model configurations. Both PSNR and SSIM metrics demonstrate that the diffusion-based models substantially
  outperform the baseline methods. Specifically, the proposed diffusion model achieved improvements of up to 1.75
  dB in PSNR and 0.0270 in SSIM compared to baselines. Between the two diffusion model
  variants, incorporating QSM and R2* maps as additional input modalities yielded nearly identical performance
  (PSNR: 29.20 vs 29.23 dB; SSIM: 0.8883 vs 0.8891), suggesting that the quantitative image similarity metrics
  alone may not fully capture the biological information preserved by these additional parametric maps. Both
  diffusion models demonstrated high fidelity to the ground truth MPRAGE images, with PSNR values exceeding 29 dB
  and SSIM scores approaching 0.89.

  \begin{table}[htbp]
      \centering
      \caption{Quantitative comparison of image synthesis performance across different models. Metrics are reported
   as Mean $\pm$ Standard Deviation over the test set (n = 98 subjects). PSNR is measured in decibels (dB), and
  SSIM ranges from 0 to 1, with higher values indicating better performance.}
      \label{tab:results_comparison}
      \begin{tabular}{lcc}
    \hline
    Model & PSNR (dB) & SSIM \\ \hline  
    U-Net & $ 28.56 \pm 2.36 $ & $ 0.8754 \pm 0.0322 $ \\ 
    Pix2Pix & $ 27.48 \pm 2.25 $ & $ 0.8621 \pm 0.0344 $ \\ 
    Diffusion (Ours) & $ 29.23 \pm 2.39 $ & $ 0.8891 \pm 0.0341 $ \\ 
    Diffusion w/ QSM R2* (Ours) & $ 29.20 \pm 2.39 $ & $ 0.8883 \pm 0.0340 $ \\ 
    \hline
\end{tabular}
  \end{table}

\begin{figure}[htbp]
      \centering
      \includegraphics[width=0.95\textwidth]{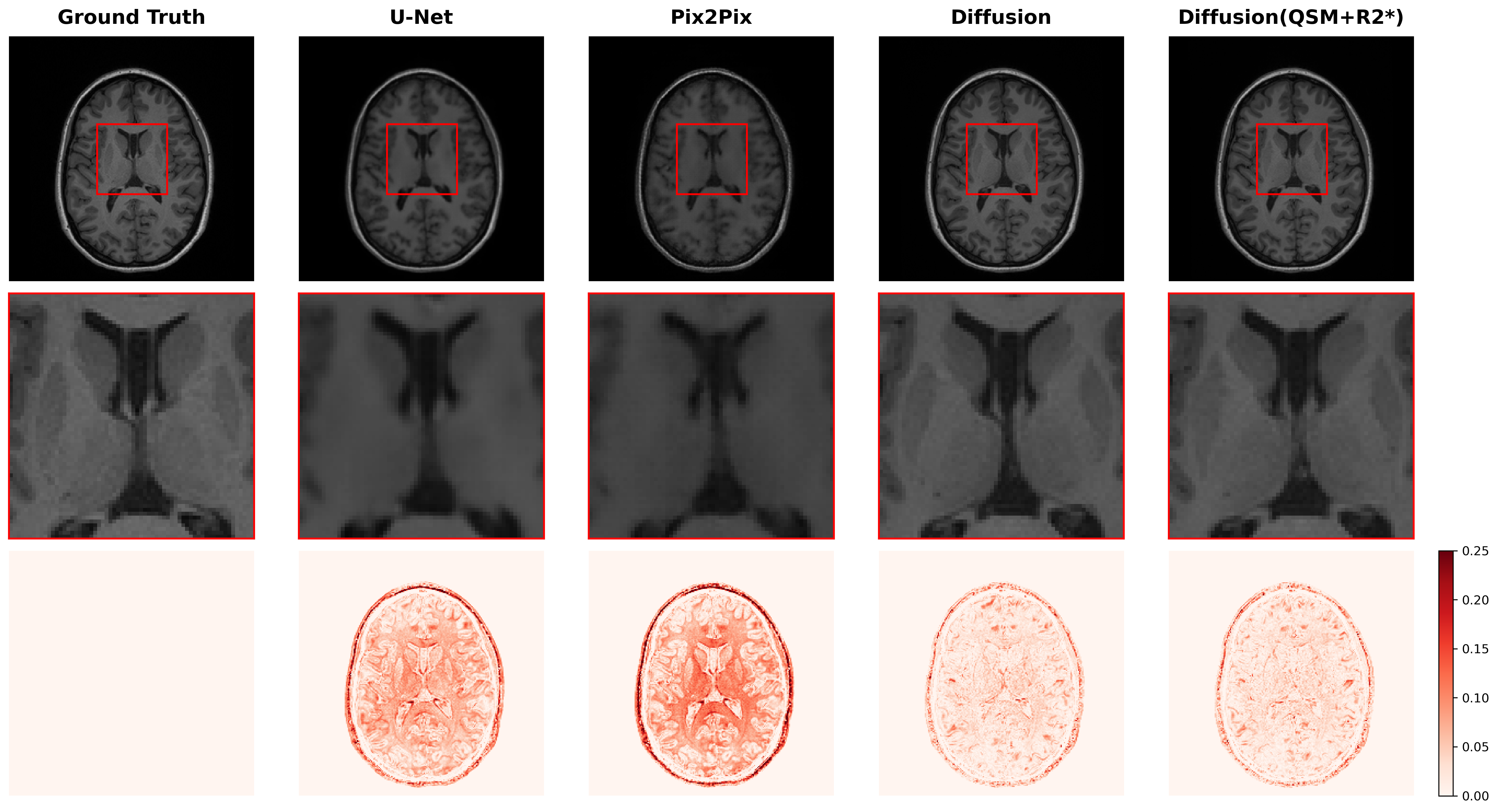}
      \caption{Qualitative comparison of synthesized T1-weighted MPRAGE images across different methods. From left
  to right: ground truth, U-Net, Pix2Pix, diffusion model, and diffusion model with QSM and R2* inputs. Top row:
  full axial slice; middle row: magnified view of the basal ganglia region (red box) showing structural details and
   tissue contrast; bottom row: absolute error maps relative to ground truth. The color bar indicates absolute
  error magnitude, with values exceeding 0.25 shown in deep red.}
      \label{fig:comparison}
  \end{figure}

  \textbf{Qualitative Analysis.} Figure~\ref{fig:comparison} provides a qualitative comparison of synthesized images across
  all methods for a representative test subject. Visual inspection reveals notable differences in image quality and
   structural fidelity. The baseline methods (U-Net and Pix2Pix) produced images with reduced sharpness and
  smoothed textures, particularly evident in the magnified basal ganglia region (middle row), where fine structural
   details and tissue boundaries appear blurred. This loss of anatomical detail is clinically significant, as
  precise delineation of brain structures is critical for accurate volumetric analysis and disease
  characterization. In contrast, both diffusion models generated images with sharper structural boundaries and
  better-preserved fine details. The absolute error maps (bottom row) demonstrate that the diffusion model with QSM
   and R2* inputs exhibits slightly reduced errors in deep gray matter regions, though the overall difference
  between the two diffusion variants remains subtle. Both diffusion models show substantially lower reconstruction
  errors compared to the baseline methods, as evidenced by the predominantly blue-green coloration in their error
  maps versus the more prominent red regions in the baseline error maps.

\begin{figure}[htbp]
      \centering
      \includegraphics[width=0.5\textwidth]{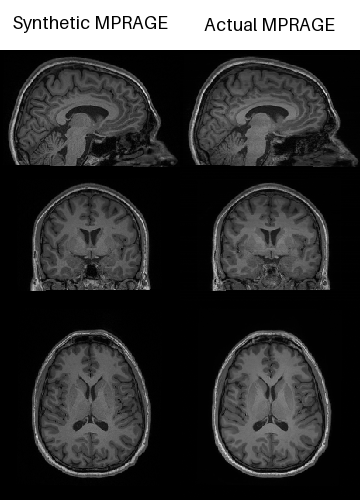}
      \caption{Representative synthesized MPRAGE images generated by the diffusion model with QSM and R2* inputs
  for one subject from the test set. Three orthogonal views are shown: sagittal (top row), coronal (middle row),
  and axial (bottom row). Left column: synthesized MPRAGE; right column: ground truth MPRAGE. The synthesized
  images demonstrate high-quality gray-white matter contrast in both cortical and subcortical regions. Despite
  being reconstructed from 2D slice-by-slice synthesis, the 3D volume exhibits smooth transitions across slices
  without visible discontinuities.}
      \label{fig:3d_reconstruction}
  \end{figure}

  Figure \ref{fig:3d_reconstruction} illustrates the 3D reconstruction quality of the diffusion model with QSM and R2* inputs across three orthogonal planes for a representative test subject. The synthesized MPRAGE images exhibit excellent gray-white matter contrast comparable to the ground truth across both cortical and subcortical regions. Although the synthesis was performed on a slice-by-slice basis in 2D, the reconstructed 3D volume demonstrates smooth inter-slice continuity without visible discontinuities or artifacts, likely attributed to the strong spatial guidance provided by the consistent 3D anatomical structure preserved in the input mGRE sequences. This volumetric consistency is essential for downstream volumetric analysis and segmentation tasks.
  
Since T1w MPRAGE images are primarily acquired for anatomical segmentation in clinical practice, we evaluated the
   functional utility of synthesized images by assessing their performance in automated brain tissue segmentation.
  This analysis provides a clinically relevant measure of whether the synthesized images can preserve the
  anatomical information necessary for accurate morphometric analysis.

\subsection{Segmentation-Based Functional Assessment}

\begin{figure*}[htbp]
      \centering
      \includegraphics[width=\textwidth]{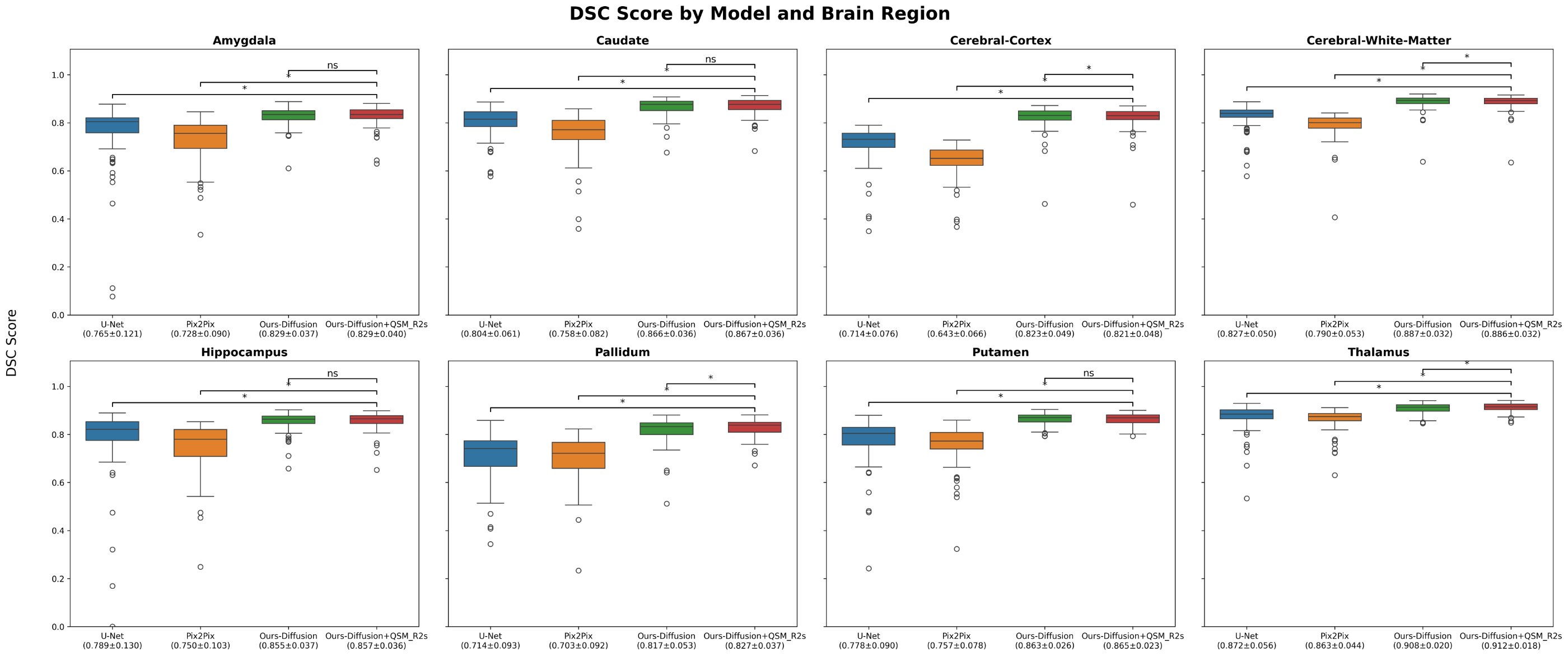}
      \caption{Segmentation accuracy comparison across eight brain regions measured by Dice Similarity Coefficient
  (DSC). Boxplots display DSC distributions for U-Net, Pix2Pix, our diffusion model, and our diffusion model with
  QSM and R2* inputs over the test set (n = 98 subjects). Each box represents the interquartile range (IQR), with
  the central line indicating the median. Whiskers extend to 1.5$\times$IQR. Statistical significance between the
  diffusion model and diffusion model with QSM/R2* is indicated by brackets: * denotes p < 0.05; ns indicates
  non-significant differences. Both diffusion models demonstrate superior performance compared to baseline methods
  across all regions.}
      \label{fig:dsc_boxplots}
  \end{figure*}

  \textbf{Dice Similarity Coefficient Analysis.} Figure~\ref{fig:dsc_boxplots} presents the distribution of DSC values
  across eight representative brain regions for all four models evaluated on the test set (n = 98 subjects). Both
  diffusion-based models substantially outperformed the baseline methods across all regions. Notably, in the four
  subcortical regions where QSM and R2* information was spatially localized (caudate, pallidum,
  putamen, and thalamus), the incorporation of these parametric maps consistently improved segmentation accuracy.
  The improvements were statistically significant in the pallidum and thalamus (p < 0.05), regions known for
   their high iron content and strong susceptibility contrast. In the cerebral cortex and white matter, regions
  where QSM and R2* provide minimal additional information, the diffusion model with QSM/R2* showed a statistically
   significant but clinically negligible decrease in DSC compared to the diffusion-only model (difference < 0.02),
  likely reflecting the increased model complexity without corresponding informational benefit in these regions.
  For the amygdala and hippocampus, where QSM and R2* maps were not included in the basal ganglia mask, no
  significant improvement was observed with the addition of these parametric inputs, as expected.

\begin{figure*}[htbp]
      \centering
      \includegraphics[width=\textwidth]{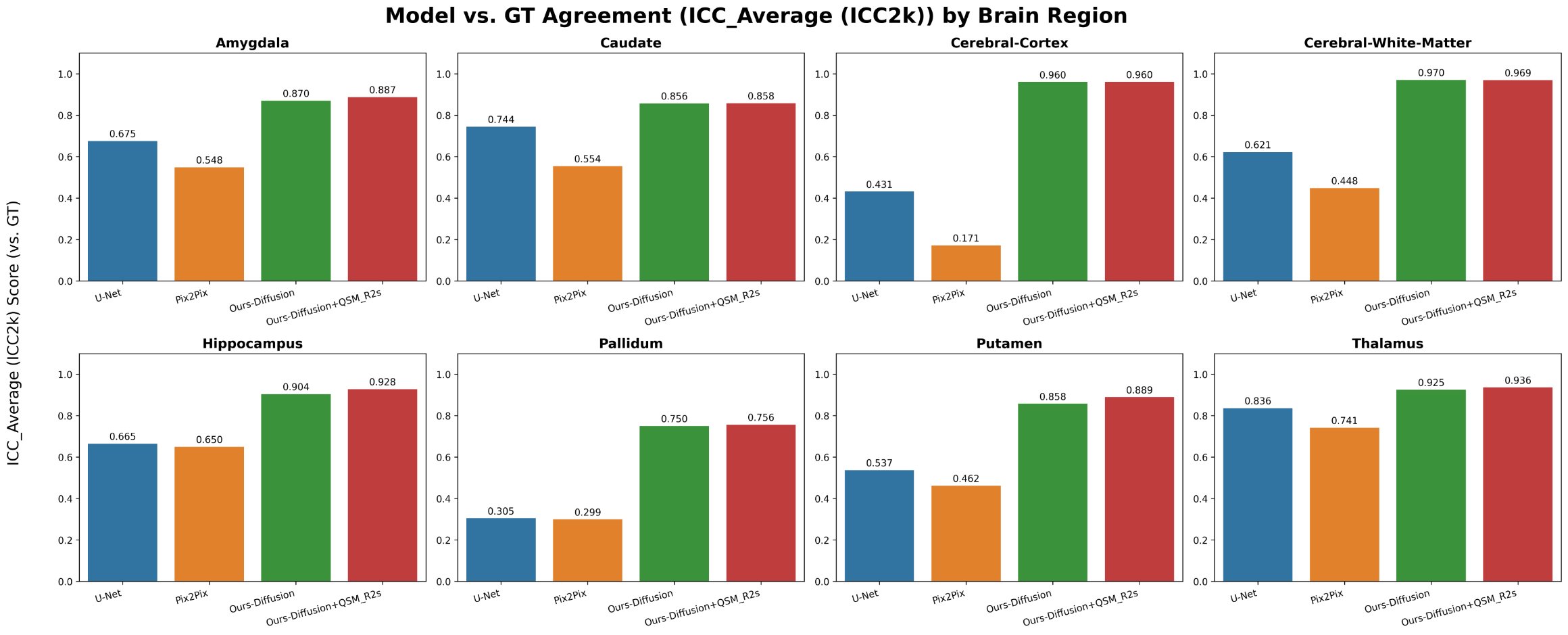}
      \caption{Volumetric measurement reliability assessed by intraclass correlation coefficient (ICC2k) across
  eight brain regions. Bar plots compare ICC values for U-Net, Pix2Pix, diffusion model, and diffusion model with
  QSM and R2* inputs. Both diffusion-based models achieve substantially higher agreement
  with ground truth MPRAGE-derived volumes compared to baseline methods, with the addition of QSM and R2* providing
   comparable or improved consistency, particularly in subcortical regions with high iron content.}
      \label{fig:icc_scores}
  \end{figure*}

  \textbf{Volumetric Reliability Analysis.} To quantify the reliability of volumetric measurements derived from synthesized
  images, we computed the intraclass correlation coefficient (ICC2k) between model-derived and ground truth volumes
   for the same eight brain regions. Figure~\ref{fig:icc_scores} shows that both diffusion models achieved
  substantially higher agreement with ground truth segmentations compared to baseline methods, with all ICC values
  exceeding the threshold of 0.75, indicating excellent reliability\cite{koo2016guideline}. Consistent with the DSC results, the four
  subcortical regions with incorporated QSM and R2* information (caudate, pallidum, putamen, and
  thalamus) exhibited improved ICC scores in the diffusion model with parametric inputs compared to the
  diffusion-only model. In cerebral cortex and white matter, ICC values remained nearly identical between the two
  diffusion variants, as expected given these regions were outside the basal ganglia mask. Interestingly, for the
  amygdala and hippocampus, the diffusion model with QSM/R2* showed marginally higher ICC values despite these
  regions not receiving direct parametric input, possibly reflecting improved overall model regularization or
  indirect benefits from the multi-modal training.

\begin{figure*}[htbp]
      \centering
      \includegraphics[width=0.8\textwidth]{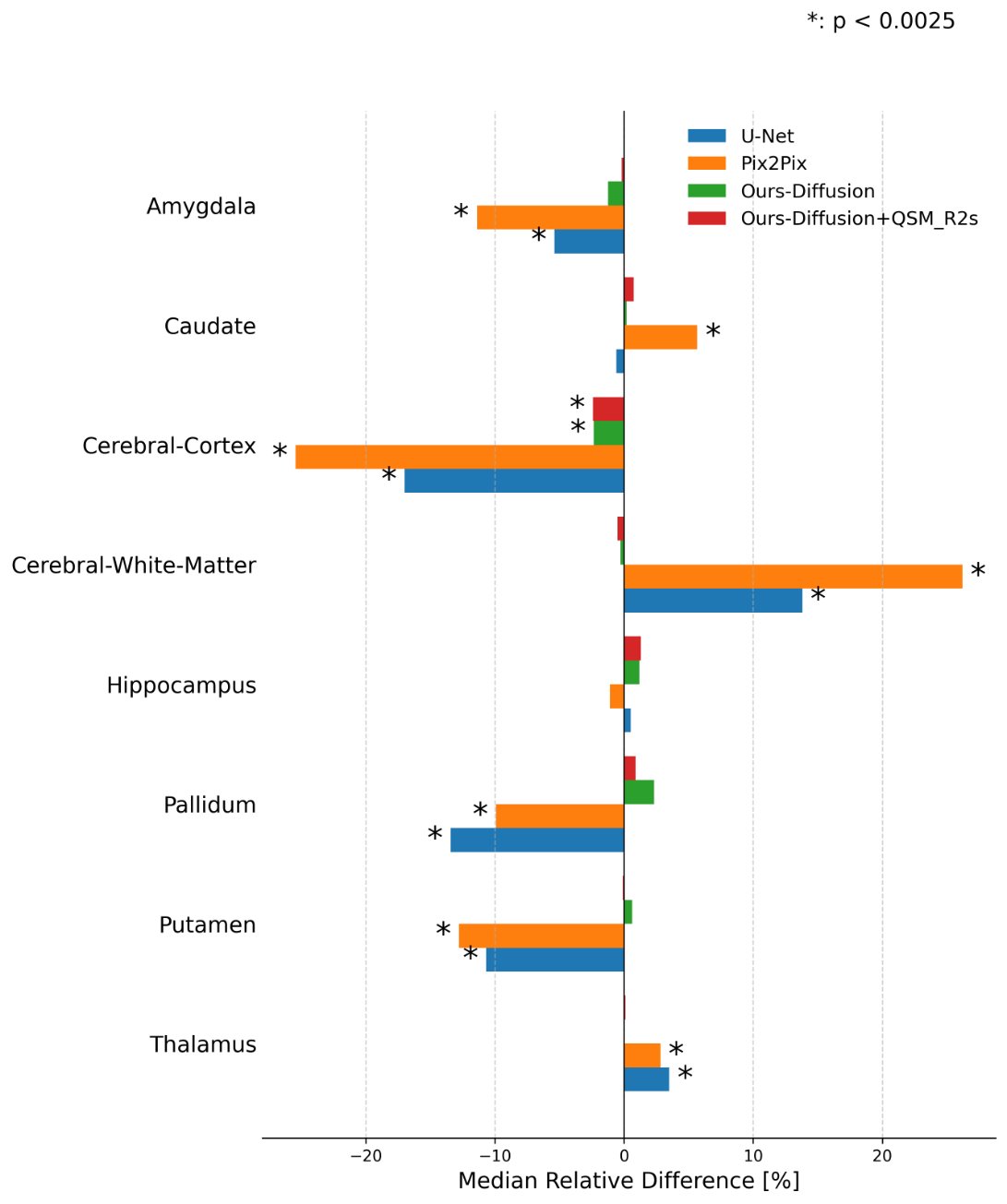}
      \caption{Median relative volumetric differences compared to ground truth across eight brain regions. Boxplots
   show the distribution of relative volume errors (computed as $\frac{V_{\text{model}} -
  V_{\text{GT}}}{V_{\text{GT}}} \times 100$) for each model. The horizontal dashed line at 0 indicates perfect
  agreement. Statistical significance is indicated by asterisks (: p < 0.0025 after Bonferroni correction). Both 
  diffusion models demonstrate superior volumetric accuracy with median differences close to zero in most regions, 
  substantially outperforming U-Net and Pix2Pix baselines.}
      \label{fig:volume_differences}
  \end{figure*}

  \textbf{Volumetric Accuracy Assessment.} Figure~\ref{fig:volume_differences} illustrates the median relative volumetric
  differences between model-derived and ground truth segmentations across all eight regions. Statistical
  significance was assessed using Wilcoxon signed-rank tests with Bonferroni correction (p < 0.0025). The results
  demonstrate that diffusion models achieved significantly better volumetric accuracy compared to baseline methods.
   Both diffusion models produced volumes that were nearly identical to ground truth measurements in most regions,
  although all four models exhibited statistically significant differences in cortical volume estimates, likely
  reflecting the inherent challenges in accurately delineating the highly convoluted cortical ribbon. Nevertheless,
   even in the cortex, diffusion models substantially outperformed baselines, with median relative differences
  approaching zero. The two diffusion model variants produced highly comparable volumetric estimates across all
  regions, with differences between them being minimal and clinically negligible.

\subsection{Biological Plausibility Assessment}

\begin{table*}[htbp]
  \centering
  \caption{Comparison of regression statistics between ground truth ($D_{\text{GT}}$) and generated ($D_{\text{GEN}}$) data across nine ROIs. The metrics include adjusted $R^2$, regression coefficient for age ($\beta_{\text{Age}}$), Cohen's $f$ for aging effect size, regression coefficient for sex ($\beta_{\text{Sex}}$), and Cohen's $d$ for sex effect size. Note that volumetric ROIs are measured in $\text{mm}^3$, and cortical thickness is measured in mm.}
  \label{tab:bio_results}
  \resizebox{\textwidth}{!}{
  \begin{tabular}{llccccccc}
    \toprule
    \multirow{2}{*}{\textbf{ROI}} & \multirow{2}{*}{\textbf{Source}} & \multirow{2}{*}{\textbf{Adj. $R^2$}} & \multicolumn{3}{c}{\textbf{Age Effect}} & \multicolumn{3}{c}{\textbf{Sex Effect}} \\
    \cmidrule(lr){4-6} \cmidrule(lr){7-9}
    & & & $\beta_{\text{Age}}$ & $p_{\text{Age}}$ & $f_{\text{Age}}$ & $\beta_{\text{Sex}}$ & $p_{\text{Sex}}$ & $d_{\text{Sex}}$ \\
    \midrule
    
    \multirow{2}{*}{Amygdala} 
    & GT  & 0.2679 & -6.182 & $7.89 \times 10^{-4}$ & 0.3579 & 309.2 & $5.21 \times 10^{-4}$ & 0.7413 \\
    & GEN & 0.2477 & -5.781 & 0.0016 & 0.3361 & 220.7 & 0.0104 & 0.5391 \\
    \midrule
    
    \multirow{2}{*}{Caudate} 
    & GT  & 0.2561 & -12.81 & 0.0031 & 0.3137 & 6.314 & 0.9753 & 0.0064 \\
    & GEN & 0.3331 & -9.946 & 0.0103 & 0.2701 & 116.9 & 0.5194 & 0.1334 \\
    \midrule
    
    \multirow{2}{*}{Hippocampus} 
    & GT  & 0.3999 & -14.66 & $4.15 \times 10^{-5}$ & 0.4436 & 486.1 & 0.0040 & 0.6093 \\
    & GEN & 0.4695 & -11.81 & $7.25 \times 10^{-4}$ & 0.3605 & 531.3 & 0.0013 & 0.6817 \\
    \midrule
    
    \multirow{2}{*}{Pallidum} 
    & GT  & 0.3946 & -7.836 & $8.83 \times 10^{-5}$ & 0.4227 & 83.97 & 0.3655 & 0.1876 \\
    & GEN & 0.2483 & -5.639 & 0.0171 & 0.2505 & 249.3 & 0.0264 & 0.4655 \\
    \midrule
    
    \multirow{2}{*}{Putamen} 
    & GT  & 0.5536 & -29.02 & $3.68 \times 10^{-11}$ & 0.7726 & 424.1 & 0.0257 & 0.4676 \\
    & GEN & 0.4641 & -24.79 & $4.16 \times 10^{-7}$ & 0.5614 & 395.2 & 0.0715 & 0.3761 \\
    \midrule
    
    \multirow{2}{*}{Thalamus} 
    & GT  & 0.6234 & -51.36 & $2.91 \times 10^{-12}$ & 0.8271 & 1270 & $8.59 \times 10^{-5}$ & 0.8469 \\
    & GEN & 0.5199 & -48.51 & $2.90 \times 10^{-9}$ & 0.6766 & 933.3 & 0.0094 & 0.5471 \\
    \midrule

    \multirow{2}{*}{Cerebral Cortex} 
    & GT  & 0.8008 & -1293 & $5.52 \times 10^{-20}$ & 1.2038 & 16650 & 0.0025 & 0.6419 \\
    & GEN & 0.8169 & -986.4 & $4.66 \times 10^{-16}$ & 1.0118 & 17110 & $5.53 \times 10^{-4}$ & 0.7376 \\
    \midrule
    
    \multirow{2}{*}{Cerebral White Matter} 
    & GT  & 0.6543 & -616.9 & $3.39 \times 10^{-4}$ & 0.3836 & 15240 & 0.0601 & 0.3926 \\
    & GEN & 0.6630 & -560.5 & $6.32 \times 10^{-4}$ & 0.3648 & 11380 & 0.1346 & 0.3113 \\
    \midrule
    
    Mean Cortical & GT  & 0.3771 & -0.0030 & $9.44 \times 10^{-12}$ & 0.7966 & -0.0074 & 0.6584 & -0.0910 \\
    Thickness & GEN & 0.3663 & -0.0019 & $2.57 \times 10^{-11}$ & 0.7752 & -0.0011 & 0.9182 & -0.0211 \\
    
    \bottomrule
  \end{tabular}%
  }
\end{table*}

To validate whether the synthesized images preserve the biological dependencies found in real populations, we performed multiple linear regression analyses on nine ROIs. Table \ref{tab:bio_results} presents the comparison of regression statistics between ground truth ($D_{\text{GT}}$) and data synthesized by our proposed diffusion model with multi-parametric integration ($D_{\text{GEN}}$).

\textbf{Aging Effects.} 
As expected in a healthy aging population, the ground truth data exhibited widespread age-related atrophy, indicated by negative regression coefficients ($\beta_{\text{Age}}$) across most regions. Crucially, the synthesized data ($D_{\text{GEN}}$) accurately replicated these trends. For instance, in the \textit{Thalamus}, a region known to be sensitive to aging, the generated data showed a regression coefficient of $\beta_{\text{Age}} = -48.51$ (p < 0.001), which is highly consistent with the ground truth ($\beta_{\text{Age}} = -51.36$). Furthermore, the effect sizes of aging, measured by Cohen's $f$, showed remarkable alignment between datasets (e.g., Thalamus: $f_{\text{GEN}}=0.6766$ vs. $f_{\text{GT}}=0.8271$; Cerebral Cortex: $f_{\text{GEN}}=1.0118$ vs. $f_{\text{GT}}=1.2038$). This suggests that our generative model successfully captures the subtle, non-linear morphological changes associated with aging.

\textbf{Sex Differences.} 
Regarding sexual dimorphism, the synthesized data preserved the general patterns of sex-related differences observed in the real data. The Cohen's $d$ values generally aligned with the ground truth categories across most ROIs. 
Crucially, for regions with negligible or small sex effects (e.g., \textit{Caudate} and \textit{Pallidum}), the generative model correctly identified them as having lower sexual dimorphism ($d_{\text{GEN}} < 0.5$) compared to highly dimorphic regions like the \textit{Thalamus} and \textit{Amygdala}. 
Although minor variations in effect size magnitudes were observed in these low-contrast regions (e.g., Pallidum), the model successfully retained the overall biological hierarchy of sex differentiation.

\textbf{Overall Model Fit.} 
The adjusted $R^2$ values, which quantify the proportion of variance explained by the demographic variables, were highly comparable between $D_{\text{GT}}$ and $D_{\text{GEN}}$. This similarity indicates that the synthesized images do not introduce varying levels of noise that would obscure biological signals.

In summary, the high concordance of $\beta_{\text{Age}}$, Cohen's $f$, and Cohen's $d$ between the two groups demonstrates that our diffusion-based method generates strictly biologically plausible images, making them suitable for downstream clinical analysis.

\section{Conclusion}
In this study, we presented a multi-parametric MRI synthesis framework based on the Fast-DDPM architecture to generate high-contrast 3D T1w MPRAGE images directly from multi-echo GRE sequences. By extending the diffusion model to integrate iron-sensitive QSM and R2* maps as conditional inputs, our method effectively addresses the challenge of contrast ambiguity in iron-rich deep gray matter structures, which is a common limitation in conventional intensity-based synthesis.

Our comprehensive evaluation demonstrates that the proposed framework significantly outperforms established U-Net and GAN-based baselines in both perceptual image quality and volumetric segmentation accuracy. Crucially, beyond pixel-level metrics, we validated the biological plausibility of the synthesized images through rigorous statistical analysis. The high concordance of aging effects ($\beta_{\text{Age}}$, Cohen's $f$) and sexual dimorphism patterns (Cohen's $d$) between the synthesized and ground truth data confirms that our generative model preserves essential population-level morphological dependencies. This indicates that the synthesized T1w images are reliable for downstream morphometric analyses.

Clinically, our approach offers a promising pathway to streamline neuroimaging protocols by eliminating the need for separate T1w acquisitions when mGRE data is available, potentially reducing scan time and patient burden. Future work will focus on extending this framework to fully 3D architectures to better capture global anatomical context, validating the model's generalizability on multi-center datasets, and exploring its applicability in detecting pathological atrophy in neurodegenerative diseases.

% --------------------------------------
\bibliography{references}

\end{document}